\documentclass[final,3p,times,twocolumn]{elsarticle}
\usepackage{graphicx}
\usepackage{amsmath,amssymb}
\journal{Physics of the Dark Universe}

\begin{document}

\begin{frontmatter}

\title{Recovering the fundamental plane of galaxies by $f(R)$ 
gravity}

\author[a]{V. Borka Jovanovi\'{c}}
\ead{vborka@vinca.rs}
\author[b,c,d]{S. Capozziello}
\author[e]{P. Jovanovi\'{c}}
\author[a]{D. Borka}
\address[a]{Atomic Physics Laboratory (040), Vin\v{c}a 
Institute of Nuclear Sciences, University of Belgrade, P.O. Box 522, 
11001 Belgrade, Serbia}
\address[b]{Dipartimento di  Fisica, Universit\`{a} di Napoli 
''Federico II'', Complesso Universitario di Monte Sant'Angelo, 
Edificio G, Via Cinthia, I-80126 Napoli, Italy}
\address[c]{Istituto Nazionale di Fisica Nucleare (INFN) Sezione di 
Napoli, Complesso Universitario di Monte Sant'Angelo, Edificio G, 
Via Cinthia, I-80126 Napoli, Italy}
\address[d]{Gran Sasso Science Institute (INFN), Viale Francesco 
Crispi 7, I-67100 L'Aquila, Italy.}
\address[e]{Astronomical Observatory, Volgina 7, P.O. Box 74, 11060
Belgrade, Serbia}

\begin{abstract}
The fundamental plane (FP) of galaxies can be recovered in the 
framework of $f(R)$ gravity avoiding the issues related to dark 
matter to fit the observations. In particular, the power-law version 
$f(R)\propto R^n$, resulting from the existence of Noether 
symmetries for $f(R)$, is sufficient to implement the approach. In 
fact, relations between the FP parameters and the corrected 
Newtonian potential, coming from $R^n$, can be found and justified 
from a physical point of view. Specifically, we analyze the velocity 
distribution of elliptical galaxies and obtain that $r_c$, the 
scale-length depending on the gravitational system properties, is 
proportional to $r_e$, the galaxy effective radius. This fact points 
out that the gravitational corrections induced by $f(R)$ can lead 
photometry and dynamics of the system. Furthermore, the main 
byproduct of such an approach is that gravity could work in 
different ways depending on the scales of self-gravitating systems.
\end{abstract}


\end{frontmatter}

\section{Introduction}
\label{sec1}

The fundamental plane (FP) of elliptical galaxies is an empirical 
relation existing between the global properties of these galaxies. 
It was first mentioned in \cite{terl81}, where authors showed that 
normal elliptical galaxies are, at least, a two-parameter family 
(the connection between velocity dispersion and line-strength). It 
was defined and discussed in more detail in e.g. 
\cite{busa97,saul13,tara15} and references therein. The structural 
properties of dynamically hot galaxies (central velocity dispersion, 
effective surface brightness and effective radius), as well as the 
global relationships of the stellar populations to their parent 
systems, are analyzed in \cite{bend92,bend93}. These fundamental 
plane properties are combined into a sort of 3D-phase space, where 
the axes are parameters that are physically meaningful.

There are two main global properties of the FP, i.e. its {\it tilt} 
and {\it tightness}. The so-called 'tilt' of FP, with respect to the 
expected virial plane, which should contain all the considered 
self-gravitating systems, means that the coefficients of the FP 
equation differ from those predicted by the Virial Theorem: if 
written in logarithmic form, the two planes appear to be tilted by 
an angle of $\sim 15^\circ$ \cite{busa97}. By analyzing the 
properties of a sample of elliptical galaxies \cite{busa97}, it 
is possible to show that more than half of the 'tilt' of the 
fundamental plane of elliptical galaxies is accounted for by the 
non-homology in the dynamical structures of the systems. It has also 
been shown that about 30 $\%$ of the tilt can be explained by 
stellar population effects and by spatial non-homology, which is 
close to the agreed-upon modern value. There are also studies 
pointing out that the tilt is mainly driven by a mass-dependent dark 
matter (DM) fraction, such that more massive galaxies have larger DM 
fractions within effective (half-light) radius \cite{tara15}. Some 
authors find that a 'hybrid' interpretation of the FP should be 
preferred, where dynamical non-homology and DM effects seem to play 
a significant role in producing the tilt \cite{dono13}.

In this paper, we want to show that it is possible to recover the FP 
by considering corrections due to modified gravity. In particular, 
we will adopt $f(R)$ gravity which is the straightforward 
generalization of Einstein's General Relativity as soon as the 
theory is $f(R)\neq R$, that is non-linear in the Ricci scalar $R$ 
as in the Hilbert-Einstein action \cite{capo11}.

The idea that modification of gravity could be used as an 
alternative to DM was first proposed by Milgrom in form of the 
modified Newtonian dynamics (MOND) \cite{milg83}. MOND gave a number 
of successful predictions in the field of galactic dynamics, and 
achieved a  significant success in explaining the flat rotational 
curves of spiral galaxies without DM, as well as in diminishing of 
its influence in ellipticals \cite{milg03}. Besides, MOND also 
inspired the development of modified relativistic gravitation 
theories \cite[see e.g.][]{beke04}. These theories have recently 
become very popular due to the fact that allow to address DM and 
dark energy issues in a self-consistent way 
\cite{capo12,noji11,capo13,bork16} without considering new material 
ingredients non-interacting electromagnetically. Besides, they fix 
several shortcomings of the Standard Cosmological Model giving rise 
to natural inflationary scenarios \cite{star80,planck}.

Here, we shall adopt a power law form of $f(R)$ gravity, that is  
$R^n$, that a part being the very first extension of Einstein 
gravity (e.g. $n\simeq 1+\epsilon$ with $\epsilon\ll 1$ 
\cite{clif05}), has an important physical meaning being determined 
by the presence of Noether symmetries in the interaction Lagrangian 
\cite{capo07b,capo08,bern11}. In other words, Noether symmetries 
determine a further effective gravitational radius (more than the 
standard Schwarzschild radius) and the value of the slope $n$. Such 
a radius and the slope of $f(R)$ function  determine the dynamical 
structure of self-gravitating systems. If confirmed by other 
observational evidences, this approach could point out that 
gravitational interaction is not scale-invariant and results holding 
for GR in Solar System could not be valid at other scales (see 
e.g. \cite{jain10}). In this sense, instead of considering dark 
matter issues, gravitational interaction should be revised at 
different scales like those of galaxies.

The main goal of this paper is to recover the FP in the context of 
$f(R)$ gravity. Such an approach is twofold. From one side, this 
geometric view could help to interpret the DM, whose effect, also 
if observationally determined at any astrophysical scale, has not 
been revealed, up to now, at fundamental particle level neither by 
direct nor by indirect detection. On the other hand, FP could be a 
formidable test bed for alternative theories of gravity at galaxy 
scales. In fact, as already pointed out for spiral galaxies in 
\cite{capo07,frig07}, elliptical galaxies in \cite{napo12}, 
and clusters of galaxies in \cite{capo09}, alternative gravity 
could result a new paradigm in order to interpret extragalactic 
astrophysical structures. In particular, as recently reported in  
\cite{salu14}, the authors analyzed rotation curves of two 
particular galaxies showing that power-law $f(R)$ gravity fits them 
better than in presence of DM.

The paper is organized in the following way. Sec. \ref{sec1} is 
devoted to power-law $f(R)$ gravity. We show that the corrected 
gravitational potential, solution of the field equations, can be 
used to model out self-gravitating systems. FP in view of $f(R)$ 
gravity is discussed in Sec. \ref{sec3}. We show how velocity 
dispersion, effective radius and surface brightness can be recovered 
starting from the parameters of $f(R)$ gravity solutions. The 
reconstruction of FP  vs observational data, considering the $f(R)$ 
interpretation, is reported in Sec. \ref{sec4}. Conclusions are 
drawn in Sec. \ref{sec5}.

\section{The gravitational potential in $f(R)$ gravity}
\label{sec2}

Let us now  shortly review the $f(R)$ gravity theory that can be 
considered a straightforward extension of Einstein's General
Relativity. The action is \,:

\begin{equation}
\label{f(R)action} {\cal{A}} = \int{d^4x \sqrt{-g} \left [ f(R) +
{\cal{L}}_{m} \right ]}
\end{equation}
where $f(R)$ is an analytic function of the Ricci curvature scalar
 $R$ and ${\cal{L}}_m$ is the standard matter contribution.
Clearly $f(R) = R$ implies  the recovering of General Relativity.
The variation of  the action with respect to the metric  $g_{\mu
\nu}$ gives the field equations [see \cite{capo02}]\,:

\begin{eqnarray}
G_{\mu \nu} & = & \displaystyle{\frac{1}{f'(R)}}
\displaystyle{\Bigg \{ \frac{1}{2} g_{\mu \nu} \left [ f(R) - R
f'(R) \right ] + f'(R)_{; \mu \nu}} \nonumber \\ ~ & - &
\displaystyle{g_{\mu \nu} \Box{f'(R)} \Bigg \}} +
\displaystyle{\frac{T^{(m)}_{\mu \nu}}{f'(R)}} \label{eq:f-var2}
\end{eqnarray}
here $G_{\mu\nu} = R_{\mu \nu} - (R/2) g_{\mu \nu}$ is the
Einstein tensor; the prime denotes derivative with respect to
$R$. The terms ${f'(R)}_{; \mu \nu}$ and $\Box{f'(R)}$ are of 
fourth order in derivatives with respect the metric $g_{\mu \nu}$. 
The case
\begin{equation}
f(R) = f_0 R^n \label{eq: frn}
\end{equation}
is selected by the existence of Noether symmetries in the theory. 
Those symmetries are related to the presence of a conserved quantity 
in the dynamical system. Physically, such a quantity gives rise to  
further gravitational radius other than the standard Schwarzschild 
radius \cite{capo07b,capo08,bern11}. A symmetry exists for any 
value of $n$ that is a real number. In the case $n=1$, the theory 
becomes again of second order (Einstein's theory) and the further 
gravitational radius is not present: in such a case, the only 
characteristic feature is the Schwarzschild radius 
\cite{capo07b}.The constant $f_0$ is chosen gives the right 
physical dimensions.

Let us now take into account the gravitational field generated by a 
pointlike source and solve the field equations (\ref{eq:f-var2}). 
Under the hypothesis of weak gravitational fields and slow motions 
(the same holding for standard self-gravitating systems), we can
write the spacetime metric in spherical symmetry as\,:

\begin{equation}
ds^2 = A(r) dt^2 - B(r) dr^2 - r^2 d\Omega^2
\label{eq: schwartz}
\end{equation}
where $d\Omega^2 = d\theta^2 + \sin^2{\theta} d\varphi^2$ is the
line element on the unit sphere. A physically motivated hypothesis
to search for solutions is (see the discussion in \cite{capo07})

\begin{equation}
A(r) = \frac{1}{B(r)} = 1 + \frac{2 \Phi(r)}{c^2}
\label{eq:
avsphi}
\end{equation}
where $\Phi(r)$ is the gravitational potential generated by a
pointlike mass $m$ at the distance $r$. It is possible to show that 
a solution is 
\begin{equation}
\Phi(r) = - \frac{G m}{2 r} \left [ 1 + \left ( \frac{r}{r_c}
\right )^{\beta} \right ]
\label{eq: pointphi}
\end{equation}
where the gravitational potential deviates from the Newtonian due to 
the correction induced by $f(R)$ gravity. 
Here 
\begin{equation}
\beta = \frac{12n^2 - 7n - 1 - \sqrt{36n^4 + 12n^3 - 83n^2 + 50n +
1}}{6n^2 - 4n + 2}
\label{eq: bnfinal}
\end{equation}
and $r_c$ is the gravitational radius induced by the Noether 
symmetry \cite{capo07b}.

Note that, for $\beta = 0$, $n=1$, the Newtonian potential is 
exactly recovered and the metric reduces to the standard 
Schwarzschild one. On the other hand, as shown in 
\cite{capo07,frig07,salu14} by taking into account extended 
self-gravitating systems, this term allows to fit galaxy rotation 
curves without the DM contribution.

Considering the solution (\ref{eq: pointphi}) and (\ref{eq: 
bnfinal}), we can search for constraints on $n$ by imposing some 
physically motivated requirements to the modified gravitational 
potential. In the following, we will consider $\beta$ and use 
Eq.~(\ref{eq: bnfinal}) to infer $n$ from the estimated $\beta$.

A first reasonable condition is the non-divergence of the potential
at infinity, that is 

\begin{displaymath}
\lim_{r \rightarrow \infty}{\Phi(r)} = 0
\end{displaymath}
and then $\beta - 1$ cannot be positive. Furthermore we can ask for 
recovering the Newtonian potential in the Solar System. As a 
consequence, one can require $\beta - 1 > -1$  to avoid increasing 
$\Phi$ in the Solar System. This means that $r\ll r_c$ and then 
Newtonian gravity is restored. These physical constraints are 
evaded considering solutions in the range

\begin{equation}
0 < \beta < 1 \  \label{eq: brange}
\end{equation}
that, by Eq.~(\ref{eq: bnfinal}), gives $n > 1$ as lower limit on 
the slope of the gravitational Lagrangian\footnote{However, as shown 
in \cite{capo07b}, the Noether symmetry exists for any value of $n$. 
This means that the presence of symmetries selects the class of 
viable theories, in this case the power-law models, but additional  
requirements have to be imposed to achieve physical models, see also 
\cite{bern11}.}.

Other considerations are in order at this point. The parameter  
$\beta$ controls the slope of the correction term while $r_c$ is 
the scale where deviations from the Newtonian potential takes place. 
For reliable physical models, $\beta$ and $r_c$ can be determined by 
observations at galactic scales. Clearly, $\beta$ should be an 
almost universal parameter while $r_c$, as in the case of the 
Schwarzschild radius, should depend on the dynamical quantities 
(mass, effective length, etc.) of the given self-gravitating system. 

Before discussing how these considerations work for FP, it is worth 
evaluating the rotation curve for the pointlike case, i.e. the 
circular velocity $v_c(r)$ of a test particle in the potential 
generated by the point mass $m$. For a central potential, it is
$v_c^2 = r d\Phi/dr$ and then, from Eq.~(\ref{eq: pointphi}), it 
is\,:

\begin{equation}
v_c^2(r) = \frac{G m}{2 r} \left [ 1 + (1 - \beta) \left (
\frac{r}{r_c} \right )^{\beta} \right ] , 
\label{eq: vcpoint}
\end{equation}
which means that the corrected rotation curve is the sum of two 
terms: the first one is equal to the half the Newtonian curve $G m
/r$, the second is the contribution related to $f(R)$ gravity. For 
$\beta = 0$, the two terms sum up and reproduce exactly the 
Newtonian result. Furthermore, for $\beta$ in the range (\ref{eq: 
brange}), the corrected rotation curve is higher than the Newtonian 
one. It is worth noticing that observations of spiral galaxy rotation
curves point out circular velocities higher than what  predicted by
the observed luminous mass and the Newtonian potential: this result 
suggests the possibility that $f(R)$ gravitational potential can 
address the problem of galactic dynamics without additional DM (see 
\cite{capo07,frig07,salu14} for a detailed discussion). Finally, 
we can say that similar considerations work also for elliptical 
galaxies \cite{napo12} and clusters of galaxies \cite{capo09}. 
This means that the whole problem of DM could be addressed 
considering modified gravity \cite{capo12}.

\section{The $f(R)$ gravity fundamental plane}
\label{sec3}

\begin{figure}[ht!]
\centering
\includegraphics[width=0.48\textwidth]{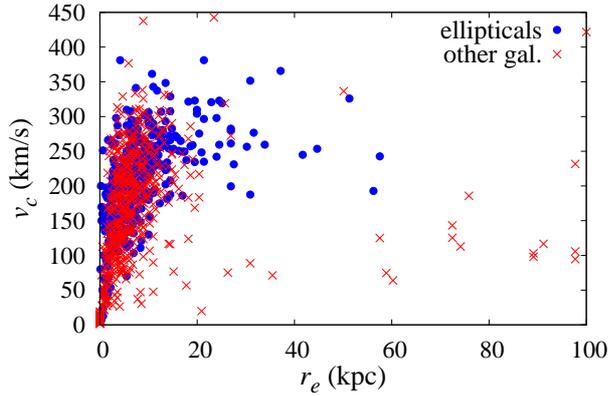}
\caption{Circular velocity $v_c$ as a function of effective radius 
$r_e$ for a sample of galaxies listed in Table 1 from 
\cite{burs97}.}
\label{fig01}
\end{figure}

\begin{figure*}[ht!]
\centering
\includegraphics[width=0.48\textwidth]{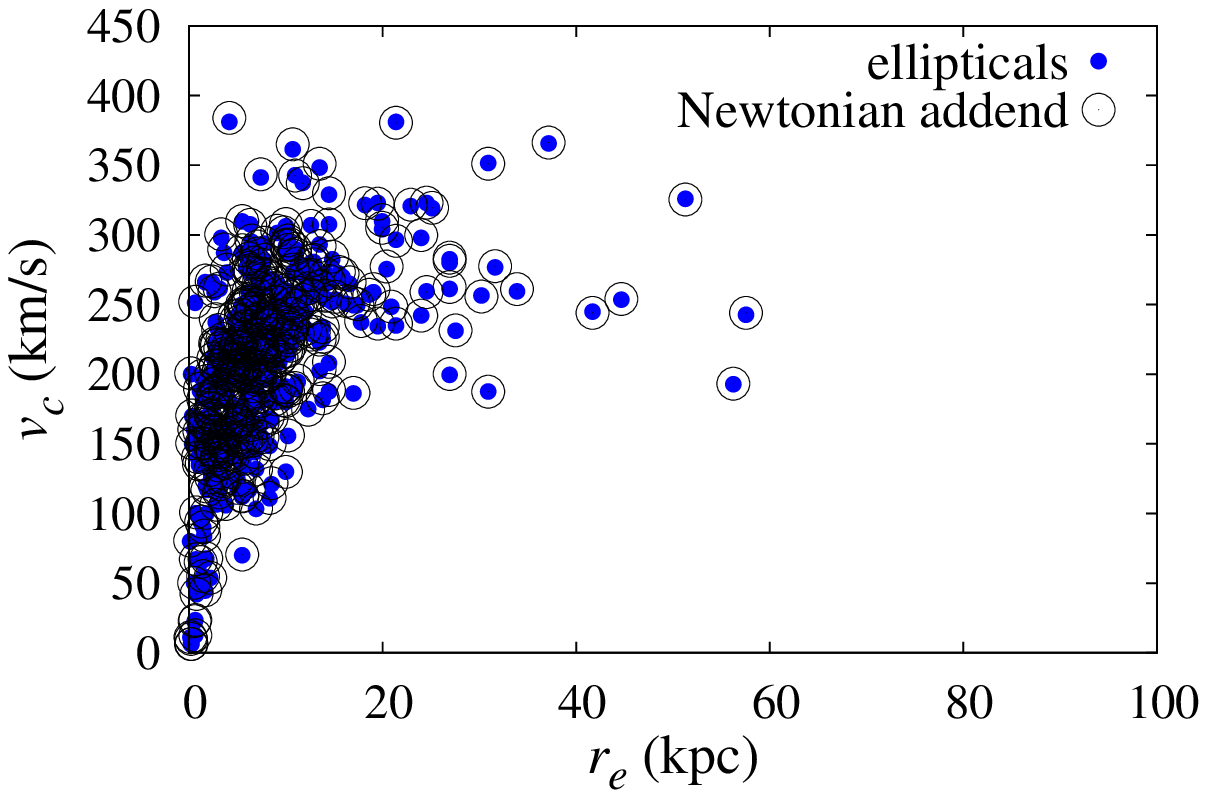}
\hspace{0.3cm}
\vspace{0.6cm}
\includegraphics[width=0.48\textwidth]{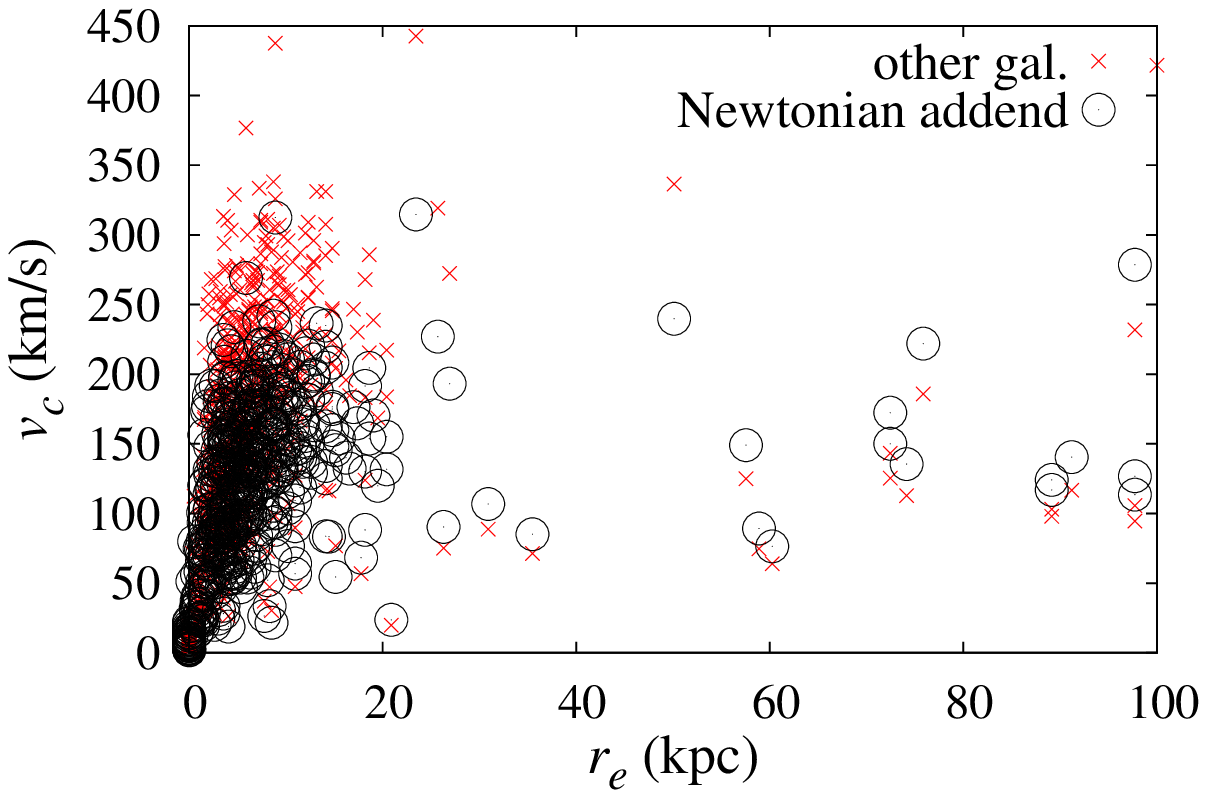}
\includegraphics[width=0.48\textwidth]{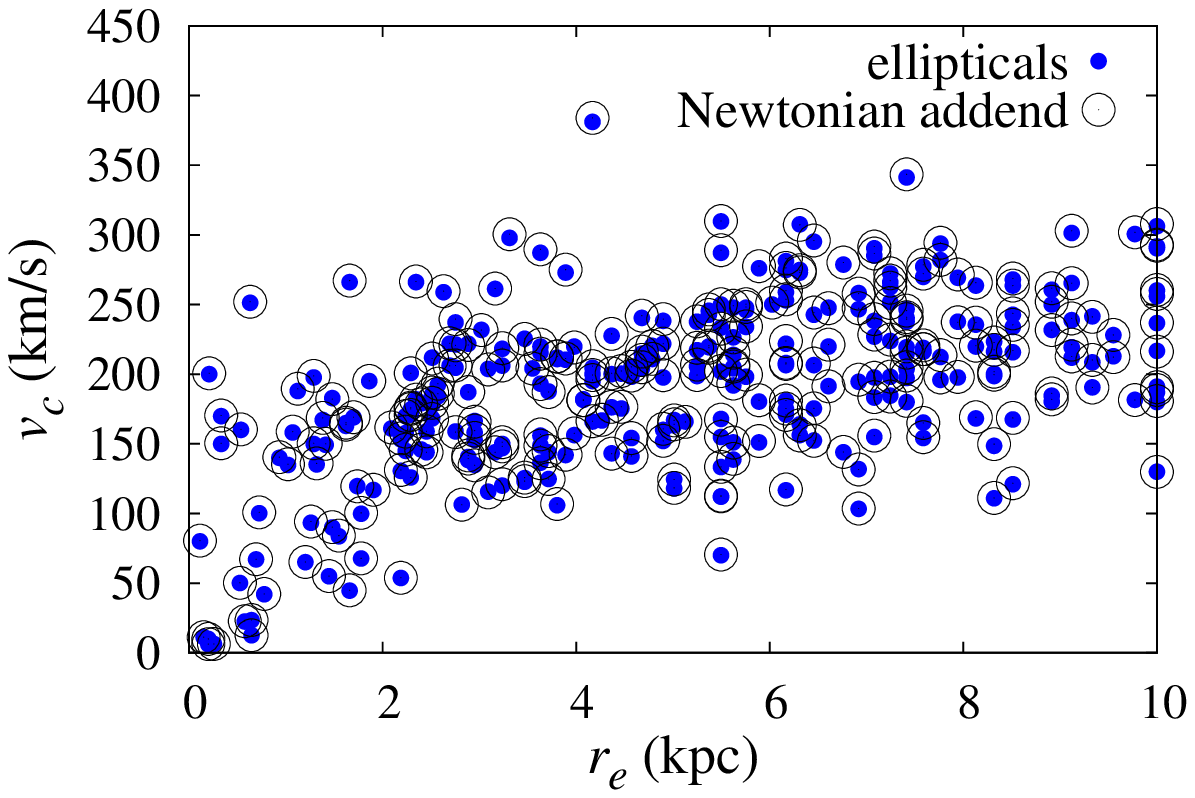}
\hspace{0.3cm}
\includegraphics[width=0.48\textwidth]{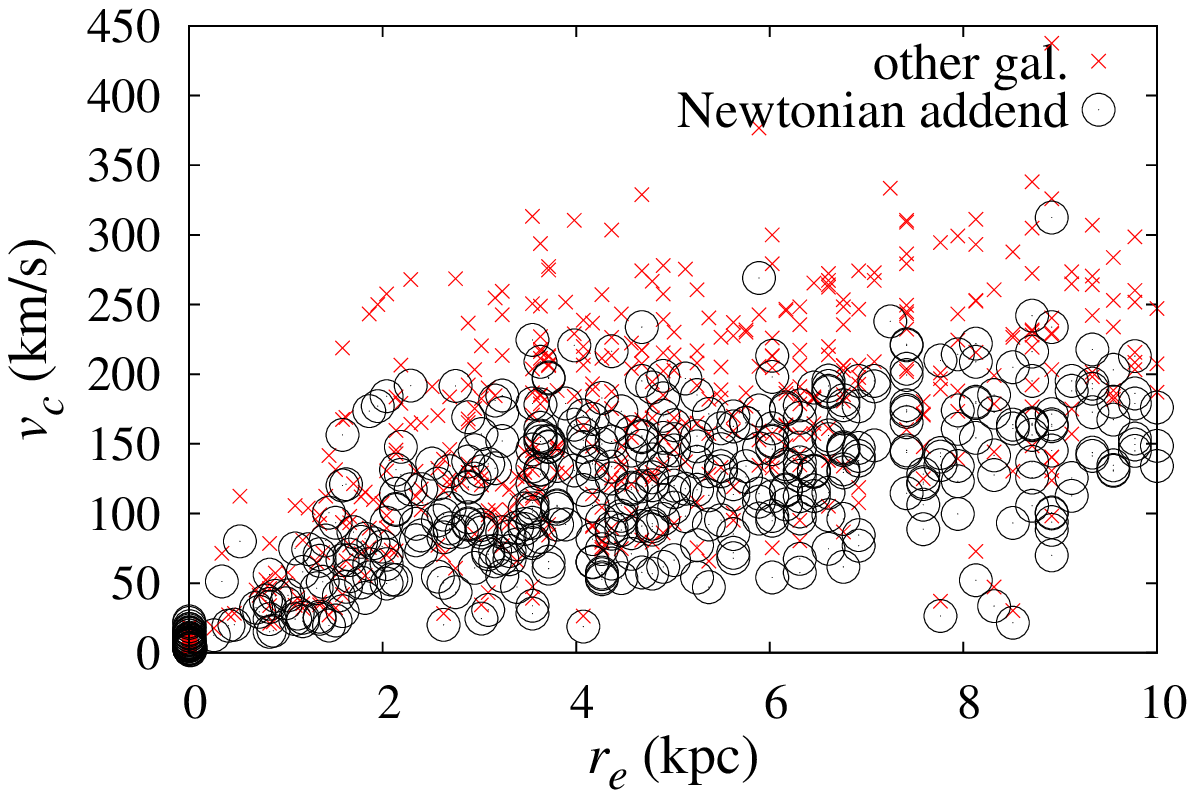}
\caption{(\textbf{left}): Circular velocity $v_c$ as a function of 
effective radius $r_e$ for elliptical galaxies (full circles) and 
their Newtonian circular velocity $v_{c,N}(r_e)$ (open circles). 
(\textbf{right}): $v_c$($r_e$) for other galaxies (crosses) and 
their Newtonian circular velocity $v_{c,N}(r_e)$ (open circles). 
Bottom panel shows a zoomed part of the figure, for $r_e$ less than 
10 kpc. Data are from \cite{burs97}.}
\label{fig02}
\end{figure*}

The FP can be written in the form \cite{busa97}:

\begin{equation}
log(r_e) = a \times log(v) + b \times log(I_e) + c,
\label{equ10}
\end{equation}

\noindent where $r_e$ is the effective radius (or half-light radius, 
within which the half of  galaxy's luminosity is contained), $v$ is 
the central velocity dispersion, and $I_e$ is the mean surface 
brightness within the effective radius. In principle, this formula 
can summarize the kinematical, morphological and photometric 
features of a given object represented as a point in the parameter 
space $\{r_e,v,I_e\}$. The velocity $v$ assumes the meaning of the 
characteristic velocity of the self-gravitating system that we are 
considering. In the case of elliptical galaxies, $v=\sigma_0$ that 
is the central velocity dispersion \cite{binn08}. This means that 
kinematics is not dominated by well defined circular velocities of 
orbiting objects (stars) as in the case of spiral galaxies (called 
"cold systems") but we need a velocity distribution as in 
thermodynamics (elliptical galaxies are called "hot systems" where 
a characteristic temperature $T$ is related to $\sigma_0$). More in 
detail, most elliptical galaxies exhibit little or no rotation. 
Their stars have random velocities along the line of sight whose 
root mean square dispersion $\sigma_0$ can be measured from the 
Doppler broadening of spectral lines \cite{binn08}.

The constant $c$ strictly depends on the morphology and the homology 
properties of the systems \cite{bend92,bend93}.

Finally, it is worth noticing that $r_e$ is typically of the order 3 
kpc for bright ellipticals and smaller for fainter galaxies 
\cite{binn08,kormendy}. It is assumed that the largest luminous mass 
fraction of the galaxy is concentrated within $r_e$ so that it can 
be considered as a typical gravitational radius, being a galaxy a 
diffuse system without a defined boundary. We want to show that the 
FP could be modulated by $f(R)$, and in such a way the issues related 
to DM could be completely reproduced. 

In this paper, we use the data given in Table I of \cite{burs97}, 
which are the result of the collected efforts of many astronomers 
over the years (as summarized in the table's explanation - available 
among the source files of its arxiv version). We want to recover 
FP using $f(R)$ gravity, which means to find connection between the 
parameters of FP equation and parameters of the potential in 
Eq.~(\ref{eq: pointphi}). In this sense, the three addends of Eq. 
(\ref{equ10}) can be connected to  the above $f(R)$ gravitational 
potential in order to find out an equation tying together 
theoretical and observational quantities. Specifically: 

\begin{itemize}
\item the addend with $r_e$ means that we need a term relating $r_e$ 
and $r_c$ in the above equation;

\item the addend with $v$  means that we need a term relating the 
velocity dispersion $\sigma_0$ and $v_{vir}$, where $v_{vir}$ is 
the virial velocity in $f(R)$;

\item addend with $I_e$ means that we need a term relating $I_e$ and 
$r_c$ (through the $r_c/r_e$ ratio).
\end{itemize}

\paragraph{\textbf{Addend with $r_e$}}

We use columns (5), (6) and (7) of Table 1 in \cite{burs97}, as 
well as the notation for circular velocity $v_c$ from that table 
that, for ellipticals, is $v_c$ = $\sigma_0$, to plot the graphic 
$v_c(r_e)$ for ellipticals and for other galaxies (see Fig. 
\ref{fig01}).

Starting from Eq.~(\ref{eq: vcpoint}), the rotation curve for 
extended systems in $R^n$ gravity, generated by spherically 
symmetric mass distribution, can be recast in the form:

\begin{equation}
v_c^2\left( r \right) = \frac{v_{c,N}^2\left( r \right)}{2} + 
\frac{r}{2} \frac{\partial {\Phi_c}}{\partial r},
\label{equ11}
\end{equation}

\noindent with $v_{c,N}^2 = \dfrac{G M(r)}{r}$ being the Newtonian 
rotation curve and $G = 4.302 \times 10^{-6} \dfrac{kpc}{M_{sun}} 
\left(\dfrac{km}{s}\right)^2$ the gravitational constant. 
$\Phi_c(r)$ is the correction term of $R^n$ gravity potential for 
spherically symmetric extended systems (see also \cite{capo07} for 
a discussion). According to Eq.~(\ref{equ11}), in the case when 
$r_c$ is proportional to $r_e$, the second addend becomes zero, so 
the circular velocity becomes exactly the Newtonian circular 
velocity $v_{c,N}$ (see Fig. \ref{fig02}). However, the velocity 
dispersion  $\sigma_0$ of elliptical galaxies is equal to their 
circular velocity $v_c$ at the effective radius $r_e$, i.e. the 
following expression holds:

\begin{equation}
\sigma_0=v_c(r_e),
\label{equ12}
\end{equation}
while this is not the case for non-ellipticals \cite{binn08}.

Let us assume now the reasonable assumption that the gravitational 
radius $r_c$  is proportional, for elliptical galaxies, to the 
effective radius:

\begin{equation}
r_c\propto r_e.
\label{equ13}
\end{equation}
This assumption is supported by the above empirical definition of 
$r_e$ which can be considered a "gravitational radius" enclosing the 
largest amount of luminous gravitating mass of the galaxy. Assuming 
the relation (\ref{equ13}) means that the theoretical gravitational 
radius $r_c$ and the observational gravitational radius $r_e$ must 
be related and eventually coincide, as in the case of Eq. 
(\ref{equ12}) where the velocity dispersion $\sigma_0$ coincides with 
the circular velocity $v_c$ at the boundary defined by $r_e$.

Using (\ref{equ13}) we have $ \xi \left( {{r_e}} 
\right) = \dfrac{{{r_c}}}{{{r_e}}} = const$, so the integrals (23) 
and (24) in \cite{capo07}, i.e. $I_1\left({r_e}\right)$ and 
$I_2\left({r_e}\right)$, do not depend on $r$. This means that the 
correction term of the gravitational potential 
$\Phi_c\left({r_e}\right)$ also does not depend on $r$. Therefore, 
it is $\dfrac{\partial {\Phi_c\left({r_e}\right)}}{\partial r}=0 
\Rightarrow v_{c,corr}=0$, and by (\ref{equ11}) and (\ref{equ12}), 
we have:

\begin{equation}
\sigma_0=\dfrac{v_{c,N}\left( r_e \right)}{\sqrt{2}},
\label{equ14}
\end{equation}

\noindent with $v_{c,N}$ being the Newtonian circular velocity and
$v_{c,corr}$ the correction term coming from  $R^n$ gravity.
In other words, under the condition (\ref{equ13}), $R^n$ gravity 
gives the same $\sigma_0$ for elliptical galaxies as in Newtonian 
case. However, this is not true for non-ellipticals, since in that 
case $v_{c,corr} \neq 0$ (see \cite{binn08} for a detailed 
discussion).

\paragraph*{\textbf{Addend with $v$}}

We can deal with objects in FP as virialized systems. In that 
sense, we can take for virial velocity in $R^n$: $v_{vir}$ = $v_c$. 
As for ellipticals (see the above discussion and explanation 
for Table 1 in \cite{burs97}) $v_c$ = $\sigma_0$, directly we have 
the connection between $\sigma_0$ and $v_{vir}$, i.e. their equality.

\paragraph*{\textbf{Addend with $I_e$}}

Correlation of this addend with $r_c$ is reflected through the 
coefficient $b$ of FP in Eq. (\ref{equ10}) or, in the other 
empirical form, $r_e \sim I_e^{\,b} v^{\,a}$ given in 
\cite{bend92,bend93}. In other words, this means that being $r_c$ 
related to $r_e$ through Eq.(\ref{equ13}), also $r_c$ is related to 
$I_e$. In general, this means that photometric quantities like $I_e$ 
are related to gravitational parameters like $r_c$.

\begin{figure}
\centering
\includegraphics[width=0.48\textwidth]{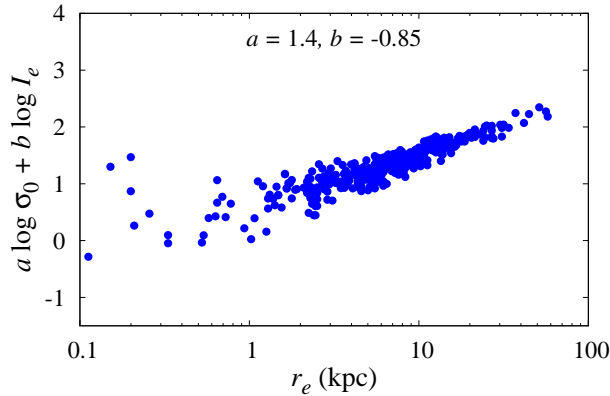}
\caption{Fundamental plane equation for observed values of 
effective radius $r_e$, velocity dispersion $\sigma_0$ and mean 
surface brightness within the effective radius $I_e$, and for 
values of $a$ and $b$ coefficients taken from \cite{bend92}: $a = 
1.4$ and $b = -0.85$.}
\label{fig03}
\end{figure}

In \cite{bend92}, it is derived the FP of elliptical galaxies, i.e. 
$a$ and $b$ are calculated as coefficients of the FP equation (using 
the Virgo Cluster elliptical galaxies as a sample). The empirical 
result is $a = 1.4, b = -0.85$. The test for our method is  
recovering this profile (see Fig. \ref{fig03}). In other words, 
starting from the gravitational potential derived from $f(R)$ 
gravity, the values of parameters $a$ and $b$ as deduced from 
observations, have to be consistently reproduced.

\section{The empirical fundamental plane reproduced by $f(R)$ 
gravity}
\label{sec4}

The theoretical circular velocity in extended spherically symmetric 
systems is calculated by using Eq. (\ref{equ11}), and taking into 
account the so called Hernquist profile for density distribution 
\cite{hern90}. This velocity is connected to $r_c$ and $\beta$ as 
given in Eq.~(\ref{eq: vcpoint}). Using these theoretical values for 
$v_c$, and the observed values for $r_e$ and $I_e$, it is possible 
to calculate a 3D fit of the FP, and as result, to obtain the 
coefficients of FP equation, that is $a$, $b$ and $c$. The method 
consists with the procedure that Eq. (\ref{equ10}) is fitted with 
\begin{equation}
\label{plane}
z(x,y) = ax + by + c\,, 
\end{equation}
where $x = \mathrm{log}(v)$, $y = \mathrm{log}(I_e)$ and $z = 
\mathrm{log}(r_e)$. Then one can calculate a table of values 
comparing the couples ($r_c$, $\beta$) vs ($a$,$b$), with  absolute 
errors for $a$, $b$, $c$, and $\chi^2$ for the fit (see Tables 
\ref{tab01} and \ref{tab02}). In such a way, we reproduce the FP 
generated by the power law $f(R)$ gravity without considering the 
presence of dark matter in galaxies. In other words, the parameters 
$r_c$ (specifically the ratio $r_c/r_e$) and $\beta$ are used to 
calculate the terms $x$ and $y$ of Eq.(\ref{plane}). Once this 
procedure is performed, the values of $a$ and $b$ that satisfy 
Eq.(\ref{equ10}) are compared with the $a$ and $b$ values 
experimentally obtained by observations \cite{bend92, bend93}. The 
final goal is to reproduce the observational FP only by parameters 
coming from the weak field limit of power-law $f(R)$ gravity.

Some important remarks are necessary at this point. Low Surface 
Brightness galaxies and galaxy clusters give a non vanishing 
beta when they are analyzed without including DM. Solar System data 
resulted with tight constraints on $\beta$, and in the case of the 
perihelion precession of Mercury gave $\beta \approx 0$, 
\cite{clif05,barr06}, practically recovering GR\footnote{In the 
formalism by \cite{clif05, barr06}, it is $\delta = 
2.7\pm4.5\times10^{-19}$, being $n=1+\delta$.}. However, 
constraints derived from  slightly larger scales, such as those 
coming from Cassini  mission, gave $\beta \approx 2 \times 10^{-6}$ 
as upper bound\footnote{In this case, it is $\delta = 
-1.1\pm1.2\times 10^{-5}$.} (see \cite{clif12} and references 
therein). Although GR is well working on these scales, some 
more recent studies indicate that it is not the case for the 
larger galactic scales where significantly larger values of 
$\beta$ can be obtained, showing that effects of $R^n$ gravity 
could start to work and should be taken into account at such scales. 
For example, as shown in \cite{bork12}, the value of this parameter 
which gives the best fit of the observed orbit of S2 star around the 
Galactic Centre is $\beta\approx 0.01$, i.e. four orders of 
magnitude larger than the previously mentioned upper bound in the 
case of Solar System. Besides, studies on Low Surface Brightness 
galaxies and galaxy clusters indicate even much larger value 
$\beta=0.817$ (see \cite{capo07,frig07,capo09}). The cases of Low 
Surface Brightness galaxies and galaxy clusters are extremely 
important in order to point out the breaking of gravity scale 
invariance at galactic scales. Specifically, the fact that 
$\beta\simeq 0$ inside the Solar System and significantly different 
from zero at galactic scales is the feature characterizing this 
issue. For example, as discussed in detail in \cite{capo07,frig07} by 
using the data reported in \cite{bosma}, the theoretical rotation 
curve of Low Surface Brightness galaxies can be modeled as a 
function of three unknown quantities, namely the stellar $M/L$ ratio 
$\Upsilon_{\star}$ and the two theory parameters $(\beta, r_c)$. 
Actually, in \cite{capo07}, the fitting parameter has been 
$\log{r_c}$ rather than $r_c$ (in kpc) since this is a more 
manageable quantity that makes it possible to explore a larger range 
of parameter space. Moreover, the gas mass fraction $f_g$, rather 
than $\Upsilon_{\star}$ is another important fitting quantity 
\cite{frig07} since the range for $f_g$ is clearly defined, while 
this is not for $\Upsilon_{\star}$. The two quantities are easily
related as follows\,:
\begin{equation}
f_g = \frac{M_{g}}{M_{g} + M_d} \iff \Upsilon_{\star} = \frac{(1 -
f_g) M_{g}}{f_g L_d}
\label{eq: fgdef}
\end{equation}
with $M_g = 1.4 M_{HI}$ the gas (HI + He) mass, $M_d =
\Upsilon_{\star} L_d$ and $L_d = 2 \pi I_0 R_d^2$ the disk total
mass and luminosity.
To constrain the parameters $(\beta, \log{r_c}, f_g)$, 
the following merit function has been minimized\,:
\begin{equation}
\chi^2({\bf p}) = \sum_{i = 1}^{N}{\left [ \frac{v_{c,th}(r_i) -
v_{c,obs}(r_i)}{\sigma_i} \right ]^2}
\label{eq: defchi}
\end{equation}
where the sum is over the $N$ observed points. While using the
smoothed data helps in better adjusting the theoretical and
observed rotation curves, the smoothing procedure implies that the
errors $\sigma_i$ on each point are non-Gaussian distributed since
they also take into account systematic misalignments between HI
and H$\alpha$ measurements and other effects leading to a
conservative overestimate of the true uncertainties (see the
discussion in \cite{bosma} for further details). As a consequence,
one does not expect that $\chi^2/dof \simeq 1$ for the best fit model
(with $dof = N - 3$ the number of degrees of freedom), but one can
still compare different models on the basis of the $\chi^2$
values. In particular, the uncertainties on the model parameters
have estimated exploring the contours of equal $\Delta \chi^2 =
\chi^2 - \chi^2_{min}$ in the parameter space. Considering the 
observational data in \cite{bosma} the best fit value for Low Surface 
Brightness galaxies is $\beta=0.817$. Similar procedures can be 
adopted for giant elliptical galaxies \cite{napo12} and galaxy 
clusters \cite{capo09}. The result is always that consistent 
deviations from Newtonian gravity (working in excellent way at Solar 
System scales) are found at galactic scales.

All these results imply that gravity, most likely, could be not a 
scale invariant interaction, and thus $\beta$ could have different 
values depending on the scale of a gravitating system. Taking this 
into account, as well as the fact that the effective galactic radii 
$r_e$ with size of $\sim 3$ kpc in average (see \cite{binn08}), have 
the significant influence on FP, it is expected that $\beta$ should 
be much larger in the case of FP than for S2 star orbit around 
Galactic Centre (which size is on the order of 1 mpc), and close to 
the values for the rotation curves of spiral galaxies. Therefore, we 
investigated the cases where $\beta$ ranges between 0.2 and 0.8 (see 
Tables 1-2 and Figs. 4-8). In summary, if further evidences will be 
found in this direction, these could indicate that the 
straightforward extrapolation of Solar System results to any 
astrophysical and cosmological scale is an approach that has to be 
carefully revised.

\begin{figure*}[ht!]
\centering
\includegraphics[width=0.98\textwidth]{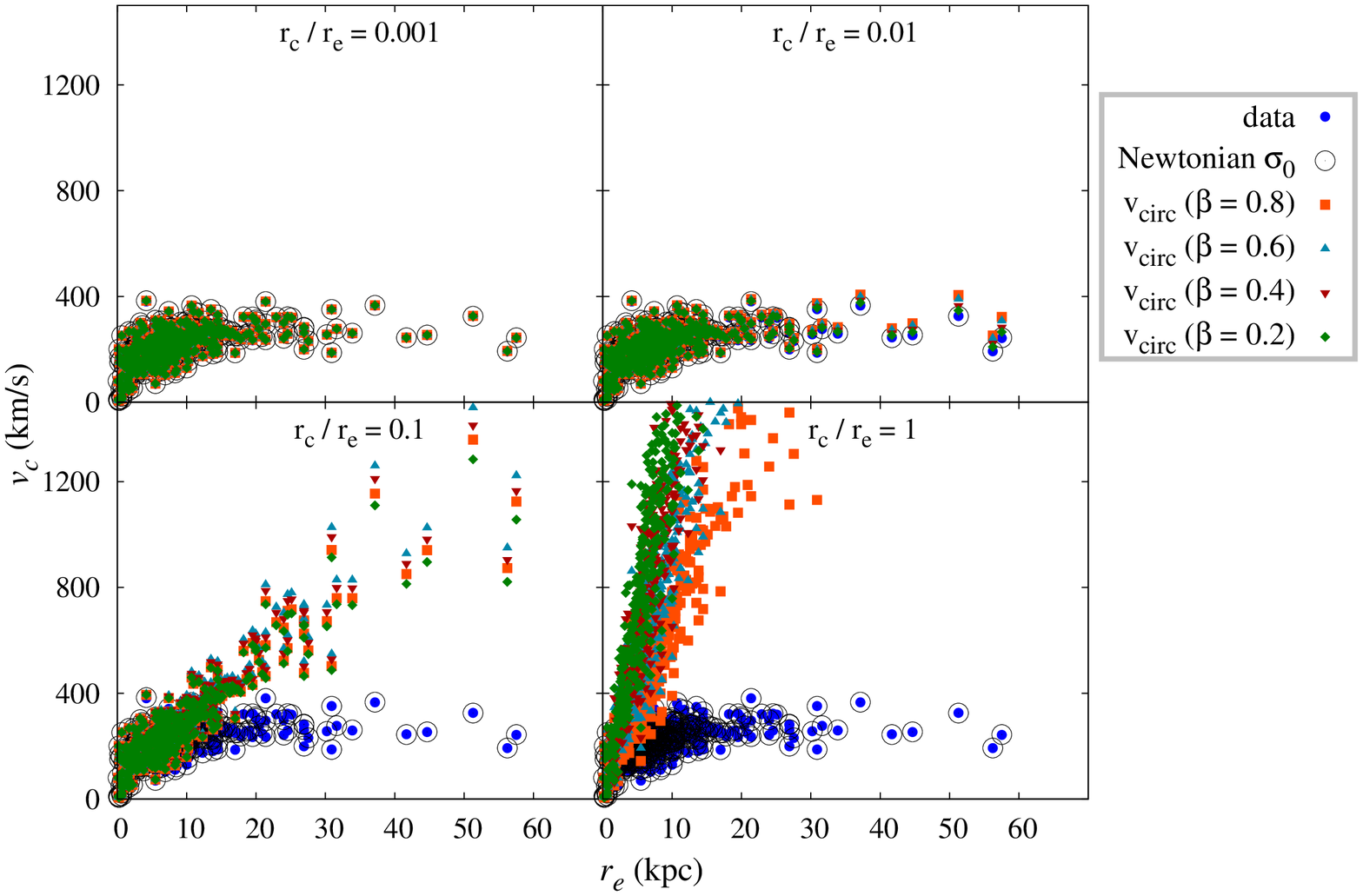}
\caption{Circular velocity $v_c$ as a function of effective radius 
$r_e$ for elliptical galaxies, for different values of ratio between 
$R^n$ gravity scale-length and effective radius $r_c/r_e$: 0.001, 
0.01, 0.1 and 1. For every of these ratios, the four values of 
$\beta$ are presented: 0.2, 0.4, 0.6 and 0.8. Notice: all $v_c$ 
values we calculated, except for blue full circles which are data 
given in Table 1 from \cite{burs97}.}
\label{fig04}
\end{figure*}

\begin{figure*}[ht!]
\centering
\includegraphics[width=0.85\textwidth]{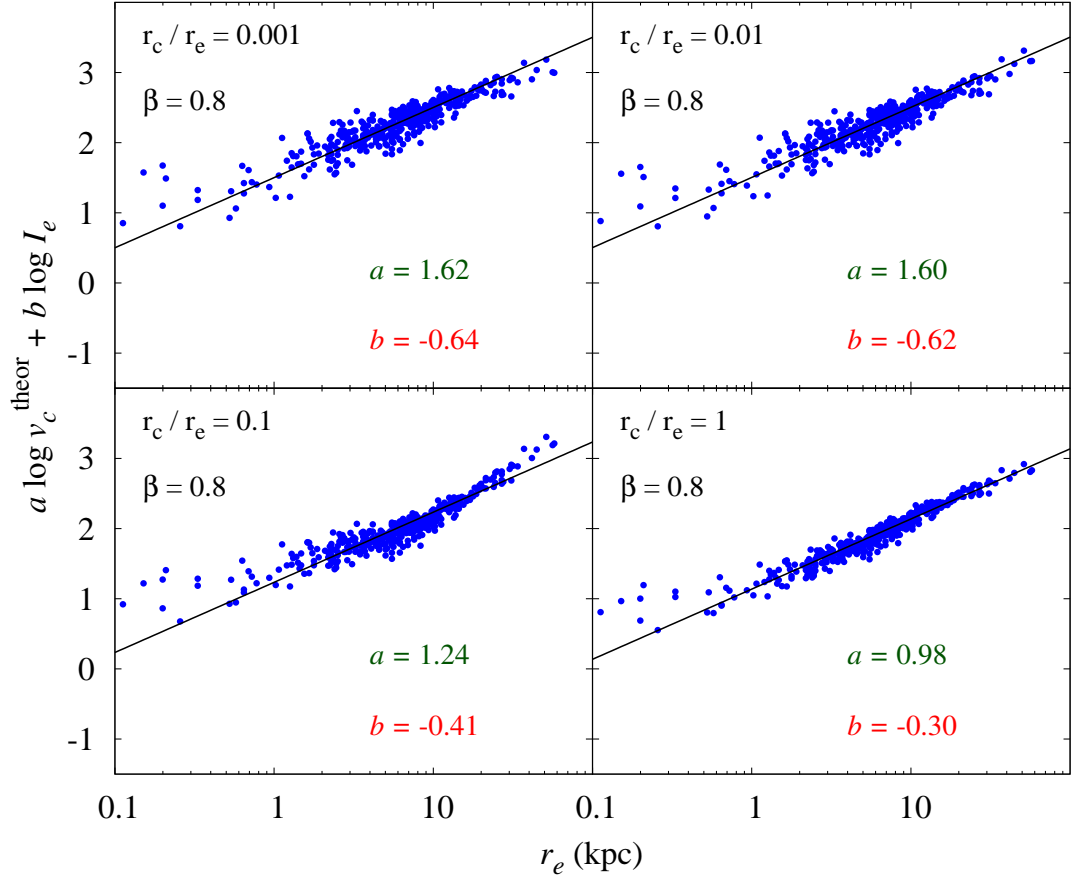}
\caption{Fundamental plane of elliptical galaxies with calculated 
circular velocity $v_c^{theor}$ and observed effective radius $r_e$ 
and mean surface brightness within the effective radius $I_e$. For 
each given pair of $R^n$ gravity parameters ($r_c$, $\beta$), 
i.e. for the certain cases $r_c / r_e$ = 0.001, 0.01, 0.1 and 1, and 
$\beta$ = 0.8, we present calculated FP coefficients ($a$, $b$). 
Black solid line is result of 3D fit of FP.}
\label{fig05}
\end{figure*}

\begin{figure*}[ht!]
\centering
\includegraphics[width=0.85\textwidth]{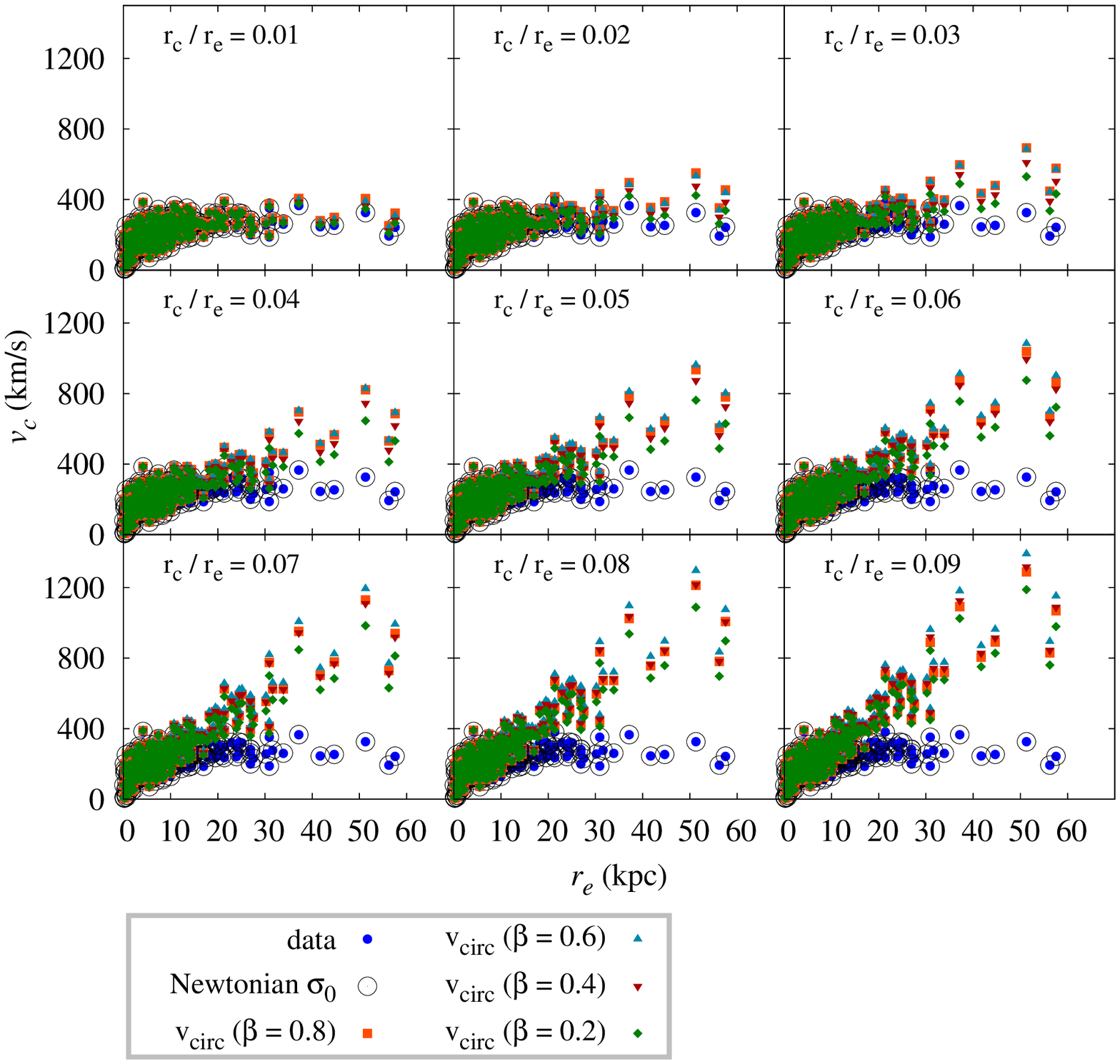}
\caption{The same as in Fig. \ref{fig04}, but for the following 
$r_c/r_e$ ratios: 0.01, 0.02, 0.03, 0.04, 0.05, 0.06, 0.07, 0.08 and 
0.09.}
\label{fig06}
\end{figure*}

\begin{figure*}[ht!]
\centering
\includegraphics[width=0.85\textwidth]{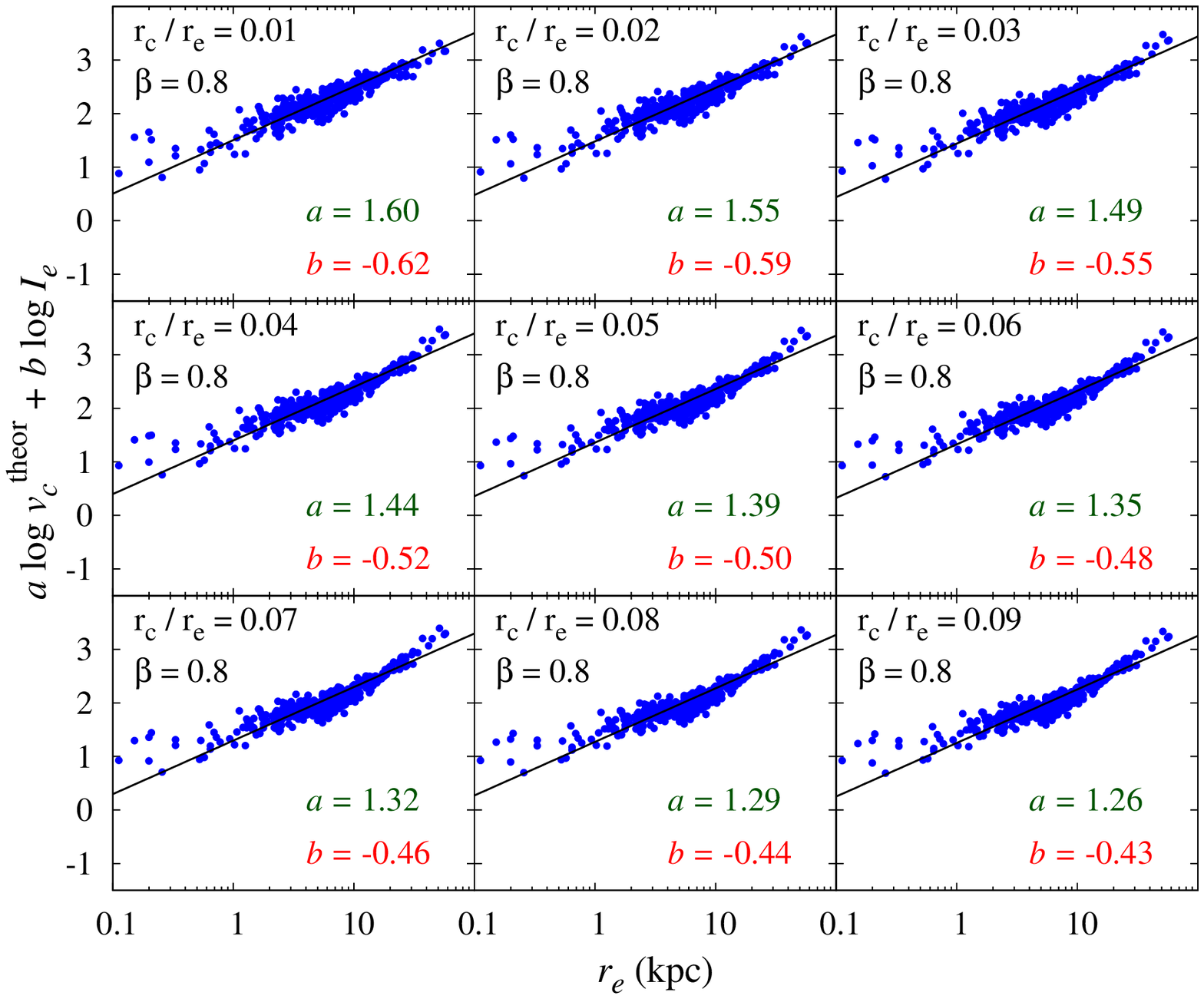}
\caption{The same as in Fig. \ref{fig05}, but for the following 
$r_c/r_e$ ratios: 0.01, 0.02, 0.03, 0.04, 0.05, 0.06, 0.07, 0.08 and 
0.09.}
\label{fig07}
\end{figure*}

\begin{figure*}[ht!]
\centering
\includegraphics[width=0.48\textwidth]{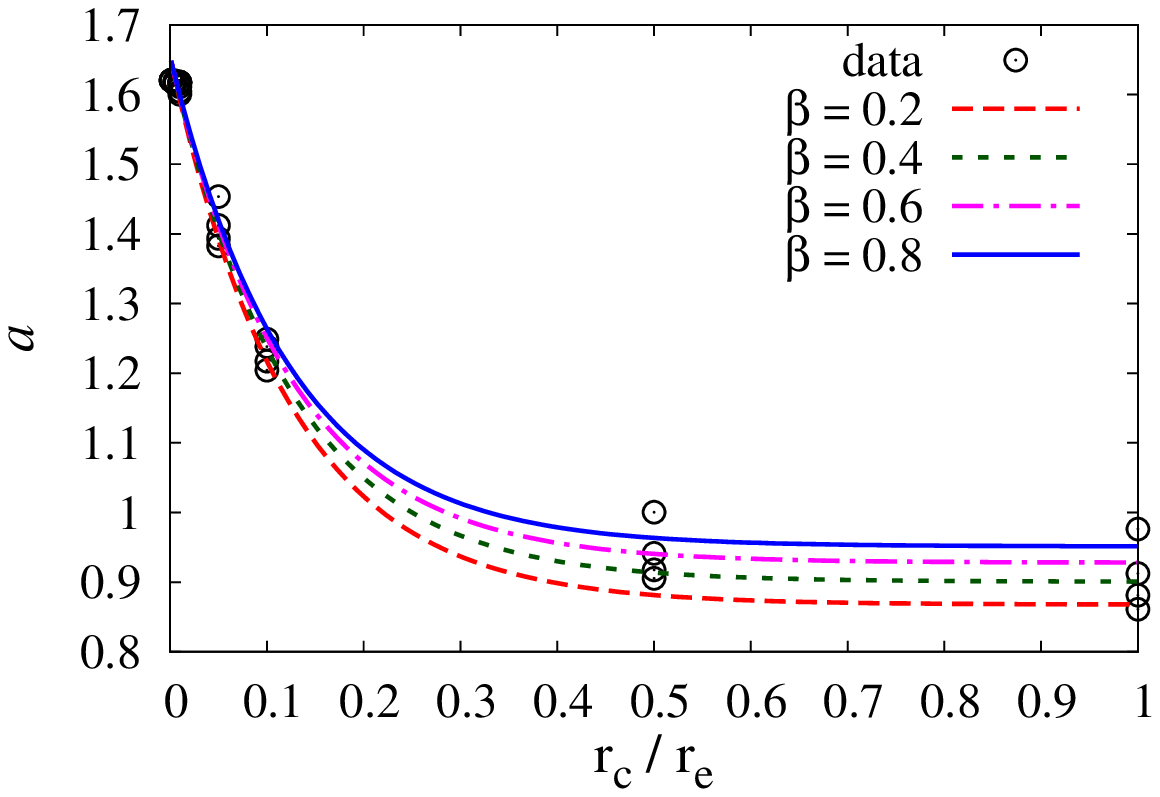}
\hfill
\vspace{0.8cm}
\includegraphics[width=0.48\textwidth]{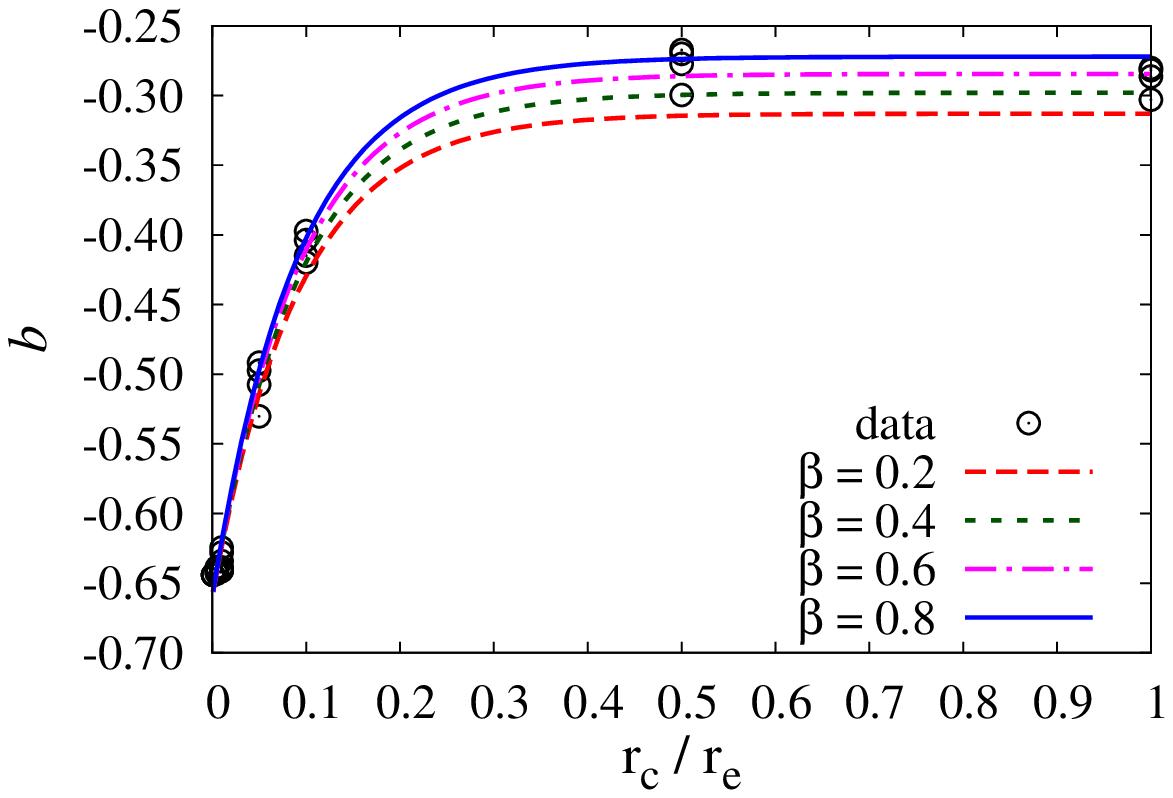}
\includegraphics[width=0.48\textwidth]{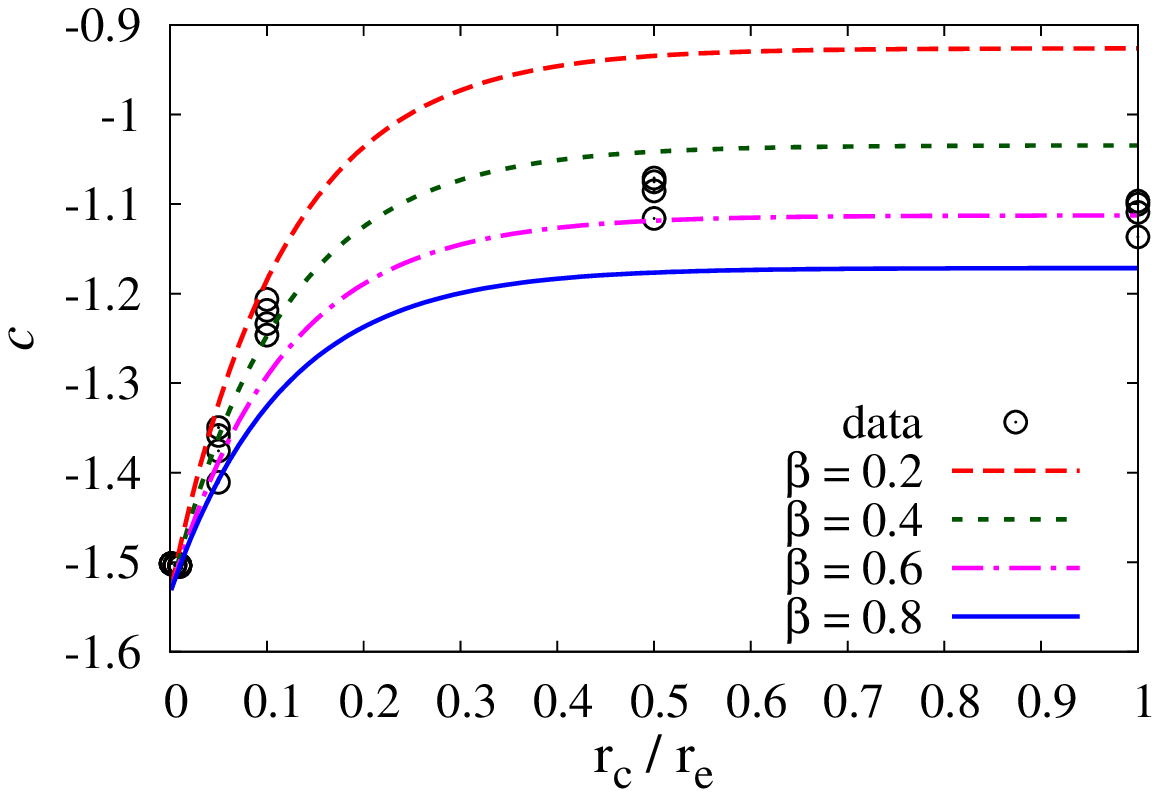}
\caption{Fits of the empirical function $f(x,\beta) = u \cdot \left( 
\dfrac{e^{-v x} - 1}{(1 + \beta)^w} + 2 \right)$ to the values of 
$a(x,\beta)$, $b(x,\beta)$ and $c(x,\beta)$ from Table \ref{tab01}, 
respectively, where $x = r_c/r_e$.}
\label{fig08}
\end{figure*}

Regarding the characteristic radius $r_c$ of $R^n$ gravity, we 
investigated its potential dependence on the effective radius $r_e$. 
For this purpose, we assumed that these two radii are proportional 
and tested their different ratios. The obtained results are 
presented in Tables \ref{tab01}-\ref{tab02} and Figs. 
\ref{fig01}-\ref{fig08}. It can be seen that for  
$r_c/r_e\approx 0.05$, the coefficient $a$ is exactly like in 
\cite{bend92}. The coefficient $b$ has similar but not exactly the 
same value. However we calculated $v_c$ only theoretically. For 
$I_e$, we are considering observed values but, in any case, the 
agreement with data is quite good (see the middle panel of Fig. 
\ref{fig07}). As discussed in Sec. 3, photometric observations gives 
$I_e$ \cite{binn08} and this allow to infer the value of $r_e$ (and 
then $r_c$ in our case). The value of $v_c$ depends on the 
self-gravitating system and, for ellipticals, from the observed 
$\sigma_0$ calculated at $r_e$. In other words, FP can be constructed 
by a set of bivariate correlations connecting some of the properties 
of (elliptical) galaxies or self-gravitating systems, including some 
characteristic radius (e.g. $r_e$), luminosity, mass, velocity 
dispersion, surface brightness, and so on. In its standard 
formulation, it is  expressed as a relation between the effective 
radius, average surface brightness and central velocity dispersion 
for normal elliptical galaxies. Any  of these three parameters can be 
estimated from the other two. The result is a plane embedded in a 
three-dimensional space \cite{bend92,bend93}.

By these considerations and results, we can state that DM effects in 
FP can be reasonably reproduced by power law $f(R)$ gravity. This is 
not surprising considering that similar results hold for single 
galaxies \cite{capo07,frig07}. Also, in Figs. \ref{fig04} and 
\ref{fig06} we present the calculated values of $v_c$ as a function 
of effective radius $r_e$ for elliptical galaxies, for different 
values of $r_c/r_e$ ratio (from 0.001 to 1), and the four values of 
$\beta$: 0.2, 0.4, 0.6 and 0.8. While in Fig. \ref{fig04} we use the 
following certain $r_c/r_e$ ratios: 0.001, 0.01, 0.1 and 1 (like 
some of them in Table \ref{tab01}), in Fig. \ref{fig06} the 
$r_c/r_e$ ratios are: 0.01, 0.02, 0.03, 0.04, 0.05, 0.06, 0.07, 0.08 
and 0.09 (like in Table \ref{tab02}). It can be noticed that 
$r_c/r_e$ has more influence on $v_c$ than $\beta$. From Fig. 
\ref{fig04}, it is noticeable that it is good agreement between 
observations and the calculations for smaller values of $r_c/r_e$ of 
the order of magnitudes $10^{-3}\div10^{-2}$, and then the agreement 
is good for all $\beta$ between 0.2 and 0.8. With increasing of that 
ratio, for better agreement, value of $\beta$ parameter has to 
decrease.

For a better insight in these results, we present the graphs with 
the FP of elliptical galaxies with calculated $v_c$ and observed 
$r_e$ and $I_e$ (Figs. \ref{fig05} and \ref{fig07}). For certain 
values of $r_c/r_e$ and $\beta$, we present calculated FP 
coefficients ($a$, $b$). Also, the 3D fit of FP is performed. From 
Fig. \ref{fig05} it can be seen that for relatively small values of 
$r_c/r_e$, the change of coefficients $a$ and $b$ with changing of 
that ratio is very small, even when $r_c/r_e$ is increased ten 
times. But for larger value of $r_c/r_e$, change of this ratio has 
greater influence on parameters $a$ and $b$. Fig. \ref{fig07} is 
useful for comparing coefficients $a$ and $b$ which we derived from 
$f(R)$ gravity, with the ones deduced from observations. When 
compare to empirical result from \cite{bend92}, it is noticeable 
that the most similar values of ($a$,$b$) correspond to $r_c/r_e$ 
interval [0.04,0.06].

\begin{table*}[ht!]
\centering
\caption{Values of $R^n$ gravity parameters ($r_c$, $\beta$) 
and the corresponding FP coefficients ($a$, $b$, $c$) calculated by 
3D fit of FP equation, with given $\chi^2$ for each fit. The chosen 
ratios between $R^n$ gravity scale-length and effective radius are 
$r_c / r_e$ = 0.001, 0.01, 0.1 and 1. For every of these ratios, the 
following four values of $\beta$ are taken: 0.2, 0.4, 0.6 and 0.8.}
\begin{tabular}{llcccr}
\noalign{\smallskip}
\hline
\noalign{\smallskip}
$r_c / r_e$ & $\beta$ & $a$ & $b$ & $c$ & $\chi^2$ \\
\noalign{\smallskip}
\hline
\noalign{\smallskip}
1 & 0.2 & 0.861 $\pm$ 0.012 & -0.280 $\pm$ 0.012 & 
-1.097 $\pm$ 0.045 & 0.00934 \\
1 & 0.4 & 0.881 $\pm$ 0.013 & -0.282 $\pm$ 0.013 & -1.100 
$\pm$ 0.047 & 0.01002 \\
1 & 0.6 & 0.912 $\pm$ 0.014 & -0.286 $\pm$ 0.013 & -1.109 
$\pm$ 0.049 & 0.01108 \\
1 & 0.8 & 0.976 $\pm$ 0.016 & -0.303 $\pm$ 0.014 & -1.137 
$\pm$ 0.054 & 0.01317 \\
&&&&&\\
0.5 & 0.2 & 0.905 $\pm$ 0.015 & -0.267 $\pm$ 0.014 & 
-1.071 $\pm$ 0.053 & 0.01305 \\
0.5 & 0.4 & 0.918 $\pm$ 0.015 & -0.270 $\pm$ 0.015 & 
-1.075 $\pm$ 0.054 & 0.01361 \\
0.5 & 0.6 & 0.941 $\pm$ 0.016 & -0.277 $\pm$ 0.015 & 
-1.085 $\pm$ 0.056 & 0.01459 \\
0.5 & 0.8 & 1.000 $\pm$ 0.019 & -0.300 $\pm$ 0.016 & 
-1.116 $\pm$ 0.061 & 0.01658 \\
&&&&&\\
0.1 & 0.2 & 1.249 $\pm$ 0.029 & -0.420 $\pm$ 0.019 & -1.246 
$\pm$ 0.079 & 0.02490 \\
0.1 & 0.4 & 1.217 $\pm$ 0.028 & -0.404 $\pm$ 0.019 & 
-1.219 $\pm$ 0.078 & 0.02491 \\
0.1 & 0.6 & 1.204 $\pm$ 0.028 & -0.397 $\pm$ 0.019 & -1.206 
$\pm$ 0.078 & 0.02499 \\
0.1 & 0.8 & 1.238 $\pm$ 0.029 & -0.415 $\pm$ 0.019 & 
-1.234 $\pm$ 0.079 & 0.02523 \\
&&&&&\\
0.05 & 0.2 & 1.454 $\pm$ 0.034 & -0.530 $\pm$ 0.020 & 
-1.411 $\pm$ 0.084 & 0.02579 \\
0.05 & 0.4 & 1.412 $\pm$ 0.034 & -0.507 $\pm$ 0.020 & 
-1.376 $\pm$ 0.083 & 0.02592 \\
0.05 & 0.6 & 1.383 $\pm$ 0.033 & -0.492 $\pm$ 0.020 & 
-1.349 $\pm$ 0.083 & 0.02606 \\
0.05 & 0.8 & 1.394 $\pm$ 0.033 & -0.498 $\pm$ 0.020 & 
-1.358 $\pm$ 0.083 & 0.02608 \\
&&&&&\\
0.01 & 0.2 & 1.616 $\pm$ 0.040 & -0.639 $\pm$ 0.021 & 
-1.504 $\pm$ 0.089 & 0.02749 \\
0.01 & 0.4 & 1.611 $\pm$ 0.039 & -0.634 $\pm$ 0.021 & 
-1.505 $\pm$ 0.088 & 0.02729 \\
0.01 & 0.6 & 1.605 $\pm$ 0.039 & -0.628 $\pm$ 0.021 & 
-1.505 $\pm$ 0.088 & 0.02706 \\
0.01 & 0.8 & 1.600 $\pm$ 0.039 & -0.624 $\pm$ 0.020 & 
-1.504 $\pm$ 0.088 & 0.02694 \\
&&&&&\\
0.005 & 0.2 & 1.620 $\pm$ 0.040 & -0.643 $\pm$ 0.021 & 
-1.502 $\pm$ 0.089 & 0.02773 \\
0.005 & 0.4 & 1.619 $\pm$ 0.040 & -0.642 $\pm$ 0.021 & 
-1.503 $\pm$ 0.089 & 0.02767 \\
0.005 & 0.6 & 1.617 $\pm$ 0.040 & -0.640 $\pm$ 0.021 & 
-1.503 $\pm$ 0.089 & 0.02757 \\
0.005 & 0.8 & 1.616 $\pm$ 0.040 & -0.639 $\pm$ 0.021 & 
-1.504 $\pm$ 0.089 & 0.02749 \\
&&&&&\\
0.001 & 0.2 & 1.620 $\pm$ 0.040 & -0.644 $\pm$ 0.021 & 
-1.501 $\pm$ 0.089 & 0.02779 \\
0.001 & 0.4 & 1.620 $\pm$ 0.040 & -0.644 $\pm$ 0.021 & 
-1.501 $\pm$ 0.089 & 0.02779 \\
0.001 & 0.6 & 1.620 $\pm$ 0.040 & -0.644 $\pm$ 0.021 & 
-1.502 $\pm$ 0.089 & 0.02778 \\
0.001 & 0.8 & 1.620 $\pm$ 0.040 & -0.644 $\pm$ 0.021 & 
-1.502 $\pm$ 0.089 & 0.02778 \\
\noalign{\smallskip}
\hline
\end{tabular}
\label{tab01}
\end{table*}

\begin{table*}[ht!]
\centering
\caption{The same as in Table \ref{tab01}, but for the following 
$r_c/r_e$ ratios: 0.01, 0.02, 0.03, 0.04, 0.05, 0.06, 0.07, 0.08 and 
0.09.}
\begin{tabular}{llcccr}
\noalign{\smallskip}
\hline
\noalign{\smallskip}
$r_c / r_e$ & $\beta$ & $a$ & $b$ & $c$ & $\chi^2$ \\
\noalign{\smallskip}
\hline
\noalign{\smallskip}
0.09 & 0.2 & 1.282 $\pm$ 0.030 & -0.437 $\pm$ 0.020 & 
-1.272 $\pm$ 0.080 & 0.02519 \\
0.09 & 0.4 & 1.247 $\pm$ 0.029 & -0.419 $\pm$ 0.019 & 
-1.241 $\pm$ 0.079 & 0.02524 \\
0.09 & 0.6 & 1.231 $\pm$ 0.029 & -0.411 $\pm$ 0.020 & 
-1.226 $\pm$ 0.079 & 0.02532 \\
0.09 & 0.8 & 1.261 $\pm$ 0.030 & -0.427 $\pm$ 0.020 & 
-1.251 $\pm$ 0.080 & 0.02549 \\
&&&&&\\
0.08 & 0.2 & 1.319 $\pm$ 0.031 & -0.457 $\pm$ 0.020 & 
-1.301 $\pm$ 0.081 & 0.02542 \\
0.08 & 0.4 & 1.281 $\pm$ 0.030 & -0.437 $\pm$ 0.020 & 
-1.268 $\pm$ 0.080 & 0.02551 \\
0.08 & 0.6 & 1.261 $\pm$ 0.030 & -0.427 $\pm$ 0.020 & 
-1.250 $\pm$ 0.080 & 0.02561 \\
0.08 & 0.8 & 1.288 $\pm$ 0.031 & -0.441 $\pm$ 0.020 & 
-1.272 $\pm$ 0.081 & 0.02571 \\
&&&&&\\
0.07 & 0.2 & 1.360 $\pm$ 0.032 & -0.478 $\pm$ 0.020 & 
-1.335 $\pm$ 0.082 & 0.02559 \\
0.07 & 0.4 & 1.319 $\pm$ 0.031 & -0.457 $\pm$ 0.031 & 
-1.299 $\pm$ 0.081 & 0.02572 \\
0.07 & 0.6 & 1.296 $\pm$ 0.031 & -0.445 $\pm$ 0.020 & 
-1.277 $\pm$ 0.080 & 0.02584 \\
0.07 & 0.8 & 1.318 $\pm$ 0.031 & -0.457 $\pm$ 0.020 & 
-1.296 $\pm$ 0.081 & 0.02589 \\
&&&&&\\
0.06 & 0.2 & 1.405 $\pm$ 0.033 & -0.503 $\pm$ 0.020 & 
-1.371 $\pm$ 0.083 & 0.02570 \\
0.06 & 0.4 & 1.363 $\pm$ 0.032 & -0.481 $\pm$ 0.020 & 
-1.335 $\pm$ 0.082 & 0.02585 \\
0.06 & 0.6 & 1.336 $\pm$ 0.032 & -0.467 $\pm$ 0.020 & 
-1.311 $\pm$ 0.082 & 0.02599 \\
0.06 & 0.8 & 1.353 $\pm$ 0.032 & -0.476 $\pm$ 0.020 & 
-1.325 $\pm$ 0.082 & 0.02601 \\
&&&&&\\
0.05 & 0.2 & 1.454 $\pm$ 0.034 & -0.530 $\pm$ 0.020 & 
-1.411 $\pm$ 0.084 & 0.02579 \\
0.05 & 0.4 & 1.412 $\pm$ 0.034 & -0.507 $\pm$ 0.020 & 
-1.376 $\pm$ 0.083 & 0.02592 \\
0.05 & 0.6 & 1.383 $\pm$ 0.033 & -0.492 $\pm$ 0.020 & 
-1.349 $\pm$ 0.083 & 0.02606 \\
0.05 & 0.8 & 1.394 $\pm$ 0.033 & -0.498 $\pm$ 0.020 & 
-1.358 $\pm$ 0.083 & 0.02608 \\
&&&&&\\
0.04 & 0.2 & 1.504 $\pm$ 0.034 & -0.560 $\pm$ 0.020 & 
-1.449 $\pm$ 0.085 & 0.02592 \\
0.04 & 0.4 & 1.466 $\pm$ 0.034 & -0.538 $\pm$ 0.020 & 
-1.420 $\pm$ 0.084 & 0.02596 \\
0.04 & 0.6 & 1.436 $\pm$ 0.034 & -0.521 $\pm$ 0.020 & 
-1.394 $\pm$ 0.083 & 0.02607 \\
0.04 & 0.8 & 1.440 $\pm$ 0.034 & -0.523 $\pm$ 0.020 & 
-1.397 $\pm$ 0.084 & 0.02609 \\
&&&&&\\
0.03 & 0.2 & 1.553 $\pm$ 0.037 & -0.590 $\pm$ 0.020 & 
-1.482 $\pm$ 0.086 & 0.02620 \\
0.03 & 0.4 & 1.523 $\pm$ 0.036 & -0.571 $\pm$ 0.020 & 
-1.462 $\pm$ 0.085 & 0.02608 \\
0.03 & 0.6 & 1.495 $\pm$ 0.036 & -0.555 $\pm$ 0.020 & 
-1.441 $\pm$ 0.085 & 0.02609 \\
0.03 & 0.8 & 1.493 $\pm$ 0.036 & -0.554 $\pm$ 0.0120 & 
-1.439 $\pm$ 0.085 & 0.02611 \\
&&&&&\\
0.02 & 0.2 & 1.593 $\pm$ 0.039 & -0.619 $\pm$ 0.021 & 
-1.502 $\pm$ 0.087 & 0.02675 \\
0.02 & 0.4 & 1.575 $\pm$ 0.038 & -0.606 $\pm$ 0.021 & 
-1.494 $\pm$ 0.087 & 0.02648 \\
0.02 & 0.6 & 1.556 $\pm$ 0.037 & -0.592 $\pm$ 0.021 & 
-1.484 $\pm$ 0.086 & 0.02631 \\
0.02 & 0.8 & 1.550 $\pm$ 0.037 & -0.588 $\pm$ 0.020 & 
-1.480 $\pm$ 0.086 & 0.02629 \\
&&&&&\\
0.01 & 0.2 & 1.616 $\pm$ 0.040 & -0.639 $\pm$ 0.021 & 
-1.504 $\pm$ 0.089 & 0.02749 \\
0.01 & 0.4 & 1.611 $\pm$ 0.039 & -0.634 $\pm$ 0.021 & 
-1.505 $\pm$ 0.088 & 0.02729 \\
0.01 & 0.6 & 1.605 $\pm$ 0.039 & -0.628 $\pm$ 0.021 & 
-1.505 $\pm$ 0.088 & 0.02706 \\
0.01 & 0.8 & 1.600 $\pm$ 0.039 & -0.624 $\pm$ 0.020 & 
-1.504 $\pm$ 0.088 & 0.02694 \\
\noalign{\smallskip}
\hline
\end{tabular}
\label{tab02}
\end{table*}

\begin{table*}[ht!]
\centering
\caption{Results of the fitting of the empirical function 
$f(x,\beta) = u \cdot \left( \dfrac{e^{-v x} - 1}{(1 + \beta)^w} + 2 
\right)$ to the values of $a(x,\beta)$, $b(x,\beta)$ and 
$c(x,\beta)$ from Table 
\ref{tab01}, where $x = r_c/r_e$.}
\begin{tabular}{lccc}
\noalign{\smallskip}
\hline
\noalign{\smallskip}
$f(x,\beta)$ & $u$ & $v$ \ & $w$ \\
\noalign{\smallskip}
\hline
\noalign{\smallskip}
$a(x,\beta)$ & \, 0.827 $\pm$ 0.004 & \, 8.1 $\pm$ 0.5 & \, 0.28 
$\pm$ 0.04 \\
$b(x,\beta)$ & -0.330 $\pm$ 0.003 & 10.9 $\pm$ 0.7 & -0.27 $\pm$ 
0.04 \\
$c(x,\beta)$ & -0.767 $\pm$ 0.009 & \, 8.5 $\pm$ 1.5 & \, 1.28 $\pm$ 
0.14 \\
\noalign{\smallskip}
\hline
\end{tabular}
\label{tab03}
\end{table*}

The next step is to find out the functional relations: $a$ = 
$a(r_c,\beta)$, $b$ = $b(r_c,\beta)$ and $c$ = $c(r_c,\beta)$. Using 
data obtained from Table \ref{tab01}, we found that $a$, $b$ and 
$c$ satisfy the following empirical dependencies on two variables: 
$f(x,\beta) = u \cdot \left( \dfrac{e^{-v x} - 1}{(1 + \beta)^w} + 2 
\right)$, with $x = r_c/r_e$. The coefficients of these empirical 
dependencies are obtained using least square fit method. We obtained 
that there is a very strong dependence on $r_c$ in the form of the 
exponential functions and a weak dependence on $\beta$ (see Table 
\ref{tab03} and Fig. \ref{fig08}). We expected the strong 
dependence on $r_c$ and a weak on $\beta$ because, like we mentioned 
in Sec. 2, $\beta$ should represent a universal constant and $r_c$ 
depends on the system's dynamical properties. From Fig. \ref{fig08} 
it can be seen that $f(x,\beta)$ describes $a$ and $b$ very well, in 
the whole range of $r_c/r_e$ from 0.001 to 1, and for all studied 
values of $\beta$.
 
As reported in Sec. 3, the value of parameter $c$ depends on the 
morphology and the homology properties of the sample of 
self-gravitating systems used to construct the FP \cite{bend92}. In 
other words, considering ellipticals, spirals, dwarfs and so on 
gives different values for $c$. Geometrically, $c$ fixes the 
intersection of FP with the axes of parameter space. Furthermore, 
morphology has a main role  in determining the DM content of systems 
so parameters like  $\beta$ and $r_c$, that correct the Newtonian 
potential, can be influenced by the observed values of $c$ 
\cite{capo07, napo12}. In particular, the observational FP reported 
in \cite{bend92,bend93} for ellipticals, is well reproduced in the 
framework of power law $f(R)$ gravity as soon as $\beta\simeq 0.8$ 
and $r_c/r_e\simeq 0.05$. In such a case, the parameters $a,b,c$ of 
Eq.(\ref{equ10}) are in excellent agreement with observations without 
considering the presence of DM.

\clearpage

\section{Conclusions}
\label{sec5}

In recent years DM as well as dark energy are become the main issues 
in cosmology and astrophysics because also if their effects are 
observed at almost every scale, their detection as new fundamental 
particles remains very elusive and complicated. In other words, also 
if the macroscopic, large scale (infrared) effects of these 
components are clearly evident, the quantum physics counterpart (new 
particles at ultraviolet scales) cannot be easily found 
out\footnote{For example, the results at CERN and other direct 
detection experiments (like LUX) indicate a formidable coherence of 
Particle Standard Model without leaving room for supersymmetric 
particles \cite{CERN}.}. This puzzling situation could be overcome 
by considering the gravitational sector and the possibility to 
extend General Relativity at large scales by retaining its positive 
results at local and Solar System scales. Addressing astrophysical 
systems and the whole Universe evolution by modified or extended 
gravity is a compelling approach with respect to introducing the {\it 
Dark Side} in the cosmological dynamics. In this perspective, FP can 
play a major role as test bed for these theories in view of 
explaining galactic structures without DM.

Specifically, FP is a formidable tool in extragalactic astrophysics 
used to study the evolution and the dynamics of self-gravitating 
systems like elliptical galaxies, bulges of spiral galaxies, 
globular clusters and so on. In some sense, the role of FP for 
galaxies could be similar to the Herzsprung-Russell diagram for 
stars: a bivariate sequence of characteristic parameters assigns the 
position and then the evolution on a sort of suitable phase-space of 
hot stellar systems. However, as reported in \cite{bend92} and 
\cite{bend93}, a quite large amount of physical processes can 
affect the shape (tightness) and the position (tilt) of FP in this 
phase-space. In particular, the presence of DM strongly affects the 
position of objects in FP determining also their evolution. 

In this paper, we consider the possibility to reproduce the FP of 
galaxies (in particular elliptical galaxies) adopting the corrected 
gravitational potential coming out as solution of power law 
$f(R)=f_0 R^n$ gravity, a straightforward extension of Einstein's 
General Relativity related to the existence of Noether symmetries 
that select the value of $n$ \cite{capo07b,capo08}. In particular, 
we have shown that $R^n$ gravity gives the same $\sigma_0$ for 
elliptical galaxies as in the Newtonian case, but for spiral 
galaxies (and other non-ellipticals) it is not the case. We obtained 
that for elliptical galaxies the characteristic radius $r_c$ is 
proportional to the effective radius $r_e$, more precise $r_c 
\approx 0.05\, r_e$ gives the best fit with data. An important point 
has to be stressed again at this point. The effective radius $r_e$, 
observationally derived from photometry, actually is a gravitational 
radius in the sense that its value is determined by the 
self-gravitating luminous matter content in the inner part of the 
elliptical galaxy. As discussed in \cite{binn08}, it can be defined 
for any spheroidal system where gravity is the dominant interaction. 
The previous assumption that $r_c\propto r_e$ is motivated by the 
fact that also $r_c$ is a gravitational radius that comes out from 
the theory. Since observational FP gives $r_c/r_e\simeq 0.05$, this 
can be considered as a sort of consistency check  that both 
(observational) $r_e$ and (theoretical) $r_c$ are gravitational 
radii.

We want to stress again that $r_c$ radius is a gravitational radius, 
and it comes out from the fact that a fourth order theory of 
gravity, like $f(R)$, gives a further gravitational radius than the 
Schwarzschild one. As shown in \cite{capo07b}, this further radius 
comes directly from the presence of a Noether symmetry. Considering 
the definition of $r_e$, we are saying that the effective radius 
(defined photometrically as the radius containing half of the 
luminosity of a galaxy) is, in some sense, led by gravity. 

In summary, our analysis shows that:
\begin{enumerate}
\item the observed FP for elliptical galaxies can be reproduced by 
$R^n$ gravity. In particular, we have shown that $R^n$ gravity gives 
the same $\sigma_0$ for elliptical galaxies as in the Newtonian case;
\item the assumption that the characteristic radius $r_c$ of 
$R^n$gravity is proportional to the effective radius $r_e$ of 
elliptical galaxies can be probed: we probed the ratio of this 
quantities and that, under this assumption, f(R) is able to 
reproduce the FP. Thus, if we take into account the definition of 
$r_e$, this indicates that the effective radius (defined 
photometrically as the radius containing half of the luminosity of a 
galaxy) is led by gravity;
\item in the range $r_c/r_e\sim 10^{-3}\div10^{-2}$, the values of 
$\beta$, which are deduced from the LSB galaxies \cite{capo07} and 
galaxy clusters \cite{capo09}, are in agreement with FP of 
elliptical galaxies. In other words, the same range of values of 
$r_c$ and $\beta$ work at galactic scales. The fact that these 
values are larger than the corresponding values obtained from the 
observational data at smaller scales, such as those from Solar 
System, may indicate that gravity is not a scale invariant 
interaction. This point deserves some comment. In \cite{frig07}, the 
analysis of Low Surface Brightness galaxies, performed in 
\cite{capo07}, has been extended to a wider class of objects 
containing also High Surface Brightness galaxies, while in  
\cite{cct} the baryonic Tully-Fisher relation has also been adopted 
besides the Low Surface galaxies sample. In the first case, the 
result was $\beta = 0.7\pm 0.25$, in the second case, it was $\beta = 
0.58\pm0.15$. Both cases are consistent with the present result 
$\beta \sim 0.8$ but the lesson is that morphology and homology 
determine corrections to the Newtonian potential. On the other hand,  
the result $\beta\simeq 0$,  from Cassini and other Solar System 
experiments, clearly means that at local scales such corrections are 
not relevant; 
\item the further gravitational radius $r_c$ is related to the 
radius of galactic bulges or the hot spherical component of 
galaxies. In the present case, it is related to $r_e$ that is the 
effective radius of ellipticals.
\end{enumerate}
In conclusion, the observed empirical FP can be reproduced by $f(R)$ 
gravity without assuming DM while the photometry of the systems is 
led by the gravitational corrections. These preliminary results can 
be improved considering detailed samples of data and different 
classes of objects. 

A final remark is now necessary. As we said above, the reported 
results seem to point out that, in absence of dark matter,  gravity 
is not a scale-invariant interaction. This means that constraints 
working at Solar System level for General Relativity could not be 
simply extrapolated at any astrophysical and cosmological scale. 
Besides, the presence of Noether symmetries could rule the strength 
and the shape of gravitational interaction at different scales. In 
our case, the couple of parameters $r_c$ and $\beta$ indicate if 
given self-gravitating systems (in this case elliptical galaxies) 
sit or not on FP. However further and detailed investigations are 
necessary to confirm this delicate statement.

\paragraph{Acknowledgments}
We wish to acknowledge the support by the Ministry of Education, 
Science and Technological Development of the Republic of Serbia, 
through the project 176003 (V.B.J., P.J. and D.B.), and Istituto 
Nazionale di Fisica Nucleare, Sezione di Napoli, Italy, iniziative 
specifiche TEONGRAV and QGSKY (S.C.). The authors also acknowledge 
the support of the Bilateral Cooperation between Serbia and Italy 
451-03-01231/2015-09/1 ''Testing Extended Theories of Gravity at 
different astrophysical scales'', and of the COST Actions MP1304
(NewCompStar) and CA15117 (CANTATA), supported by COST (European 
Cooperation in Science and Technology).

\end{document}